# Towards Scalable Subscription Aggregation and Real Time Event Matching in a Large-Scale Content-Based Network

Ruisheng Shi[*], Lina Lan[*], Peng Liu[†], Di Ao[*], Yueming Lu[*]

[*]School of Cyberspace Security
Beijing University of Posts and Telecommunications
[†]College of Information Sciences and Technology
Pennsylvania State University

## ABSTRACT

Although many scalable event matching algorithms have been proposed to achieve scalability for large-scale content-based networks, content-based publish/subscribe networks (especially for large-scale real time systems) still suffer performance deterioration when subscription scale increases. While subscription aggregation techniques can be useful to reduce the amount of subscription dissemination traffic and the subscription table size by exploiting the similarity among subscriptions, efficient subscription aggregation is not a trivial task to accomplish. Previous research works have proved that it is either a NP-Complete or a co-NP complete problem.

In this paper, we propose DLS (Discrete Label Set), a novel subscription representation model, and design algorithms to achieve the mapping from traditional Boolean predicate model to the DLS model. Based on the DLS model, we propose a subscription aggregation algorithm with $O(1)$ time complexity in most cases, and an event matching algorithm with $O(1)$ time complexity. The significant performance improvement is at the cost of memory consumption and controllable false positive rate. Our theoretical analysis shows that these algorithms are inherently scalable and can achieve real time event matching in a large-scale content-based publish/subscribe network. We discuss the tradeoff between memory, false positive rate and partition granules of content space. Experimental results show that proposed algorithms achieve expected performance. With the increasing of computer memory capacity and the dropping of memory price, more and more large-scale real time applications can benefit from our proposed DLS model, such as stock quote distribution, earthquake monitoring, and severe weather alert.

## KEYWORDS

Content-based Publish/subscribe, Event Matching, Subscription Aggregation, Subscription Subsumption Checking, Subscription Merging, Bloom filters (BF)

[*] Corresponding author:shiruisheng@bupt.edu.cn

## 1 INTRODUCTION

Information is flooding in our era. While we dislike to be bothered by useless information, on the other hand, we do not want to miss any piece of useful information and hope critical messages can arrive timely. The content-based publish/subscribe paradigm [1] enables users to express their interests flexibly by subscriptions, and receive matching events timely. To handle an enormous number of events, pub/sub overlay network can be constructed to improve system throughput [2]. Network is not the bottleneck of information processing since published messages can reach subscribers timely from any broker node in content-based publish/subscribe network.

Since no subscription shall be ignored, each broker needs to know the subscriptions of all its clients to assure that no subscribed events be dropped. Brokers have to process an enormous number of subscriptions in a large-scale content-based network, so event matching speed becomes the bottleneck. Unfortunately, this problem cannot be solved by adding more broker nodes into the overlay network. Moreover, to time critical applications, every second counts. For example, one second delay of stock trading information means loss of money, and one second delay of emergency alerts means loss of lives. Real time event matching really matters, and we must develop efficient techniques.

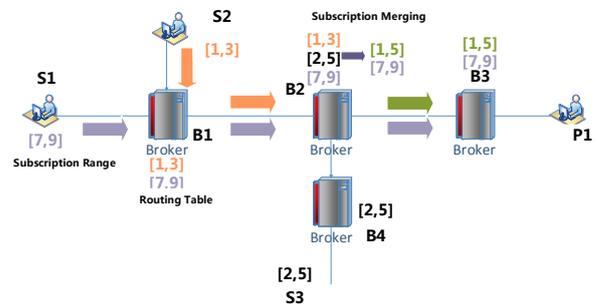

**Figure 1:** An example of Subscription Aggregation

Although many scalable event matching algorithms [5-14] have been proposed to achieve scalability for large-scale content-based networks, content-based publish/subscribe networks (especially for large-scale real time systems) still suffer performance deterioration with the increase of subscriptions. Therefore, subscription aggregation techniques have been proposed to reduce both the traffic of subscription dissemination and the subscription table size by exploiting the similarity among subscriptions. Figure 1 shows an example.

Unfortunately, subscription aggregation is not a trivial task. There are two categories of subscription aggregation techniques: subscription subsumption checking and subscription merging. The subsumption checking problem has been shown to be co-NP complete [3]; the general subscription merging problem has been proved to be NP-complete [4].

In this work, we make the following contributions:

1. We propose DLS (Discrete Label Set), a novel subscription representation model, and design algorithms to achieve the mapping from traditional Boolean predicate model to DLS model. In DLS model, the content space is partitioned into $N_r$ ranges. Each range is a subset of one subscription and is assigned a unique label. Thus, a subscription is represented as a label set. We store this label set into a Counting Bloom filters. An event is represented as a label since each event must fall into some range in the content space. This model can support subscriptions of discrete attributes.
2. We propose a concise label encoding scheme to reduce the traffic of subscription messages. This range label encoding scheme can accommodate multiple businesses that share one content-based network to avoid publishing an event multiple times.
3. Based on DLS model, we proposed an efficient subscription aggregation algorithm with average $O(1)$ time complexity and an event matching algorithm with $O(1)$ time complexity. Although the worst case of subscription aggregation algorithm has the time complexity of $O(N_r)$, it seldom occurs for most subscriptions.
4. We carry out extensive experiments to evaluate the performance of our approach.

## 2 RELATED WORK

### 2.1 Event Matching Algorithms

There has been a large body of work on content-based pub/sub systems to improve the performance of large scale event matching.

Most competitive event matching algorithms [5-11, 13, 15] have achieved $O(logN)$ time complexity. However, these algorithms still cannot achieve the goal of real time event matching when the subscriptions reach a very large scale.

The Summary Instance (SI) scheme proposed in [12] is a $O(1)$ event matching algorithm. Unfortunately, the SI scheme cannot accommodate dynamic rule set; thus, it is infeasible for subscription aggregation since subscriptions are dynamic.

[14] proposes an efficient event routing method based on Bloom filters. However, the method only supports range subscriptions.

### 2.2 Subscription Aggregation Techniques

Subscriptions typically are broadcast to all brokers. Broadcasting all subscriptions over the network significantly increases subscription dissemination traffic and subscription table sizes at nodes, which makes content matching more expensive during publication dissemination. To reduce subscription forwarding traffic and subscription table sizes at nodes, subscription aggregation techniques are used to prevent propagation of redundant subscriptions.

Specifically, researchers have proposed two kind of subscription aggregation techniques: subscription subsumption checking and subscription merging. Subscription covering is a special case of subscription subsumption. SIENA [1] is a content-based pub/sub system that uses covering-based routing, and REBECA [18] uses merging-based routing. These two techniques are both applied in PADRES [19, 20].

When a new subscription comes in, the broker needs to check if this new subscription is subsumed by the current active subscription set. If it is completely subsumed by the existing subscription set, it will not be forwarded. This technique can achieve greater efficiency in terms of reducing traffic among nodes and reducing sizes of the subscription tables at nodes.

Efficient subscription subsumption checking is not a trivial task. The subsumption checking problem where subscriptions can be represented as convex polyhedra has been shown to be co-NP complete [3].

Subscription merging is to replace a set of subscriptions with one subscription that have the same effect on publication routing. Subscription merging can further reduce the footprint of subscription table and traffic between nodes. The general efficient subscription merging problem has been proved to be NP-complete [4].

Oukselet et al. [17] present a 'Monte Carlo type' probabilistic algorithm for subsumption checking. This algorithm may result in false negatives in publication dissemination. In this algorithm it is possible that propagation of a subscription is stopped while it is not subsumed by the existing subscriptions. This may result in not delivering publications to some subscribers, which is not acceptable in critical applications like earthquake monitoring, severe weather alert. The algorithm has $O(k*m*d)$ time complexity, where $k$ is the number of subscriptions, $m$ is the number of distinct attributes (dimensions) in subscriptions, and $d$ is the number of tests performed to detect subsumption of a new subscription.

While the subsumption problem in general case is co-NP complete, if the subscriptions are d-dimensional rectangles the problem can be solved in $O(n^d)$, where n is the number of existing subscriptions, d is the number of distinct attributes (dimensions) in subscriptions. Jafarpour et al. [16] propose an approximate subscription subsumption checking approach based on negative space representation. This approximate subsumption checking algorithm reduces subscription forwarding traffic without affecting the correctness of execution. The approximate approach provides knobs to control the accuracy of subsumption checking by adjusting the required space and time.

Jafarpour et al. [15] propose the MICS (Multidimensional Indexing for Content Space) representation and processing model. In MICS, subscriptions are mapped to a single dimensional space using Hilbert space filling curve and are represented using a set of ranges. Based on this representation, subscriptions are organized in a B+ Tree. The event matching algorithm tries to find the leaf in the B+ tree that the event may fall in. A new subscription is updated into a B+ tree by a subscription aggregation algorithm. Through model mapping, the time complexity of the algorithm depends on the number of ranges in the new model.

Our proposed DLS model shares the content space partition idea with MICS. DLS adopts different mapping models, and each subscription is represented as a set of discrete labels. Routing table (i.e. subscription set) is stored into Counting Bloom filters. Thus, event matching algorithm can achieve *O(1)* time complexity at the cost of some false positives. A new subscription is updated into Counting Bloom filters by subscription aggregation algorithms. CBF can accommodate *unsubscribe* operation.

## 3 MODEL DESCRIPTION

### 3.1 Subscription and Event Model

Assume the content space consists of $d$ attributes $\{a_i\}$. All events form a $d$-dimensional space.

A publication is also called an event in the context of content-based pub/sub systems. An event is described by a set of attributes, $e = <v_1, v_2, ..., v_d>$. The attribute vector $A = \{a_1, a_2, ..., a_d\}$ is also called event schema. Here we denote the domain of attribute $a_i$ with $[l_i, u_i]$ where $l_i$ is the lower bound of the attribute domain and $u_i$ is the upper bound.

A subscription *s* is a conjunction of Boolean predicates:

$$s = \{P_1 \wedge P_2 \wedge \cdots \wedge P_d\} \quad (4)$$

Each predicate is a 4-tuple [9, 13, 15]:

$$P_i = \{attr, lowVal, highVal, type\} \quad (5)$$

where $attr$ is attribute name, $lowVal$ and $highVal$ are the bounds of attribute values, and $type$ is data types. Thus, a subscription corresponds to a $d$-dimensional rectangle in the content space. Single-sided constraints, like $\{attr, operator, Val, type\}$, are implicitly converted to range constraints by including a MAX or MIN attribute value.

The DLS model also supports subscriptions of discrete attributes, which will be discussed in Section 5.

### 3.2 DLS Model

To be able to efficiently support subscription management, we map the multi-dimensional content space to a set of labeled ranges.

Each dimension domain is divided into $g_i$ intervals where $i$ is from *1* to *d*. Thus, the content space is partitioned into $N_r$ ranges where $N_r = \prod_{i=1}^{d} g_i$ is the number of ranges in DLS model. For the convenience of analysis, we can assume $g_i = 2^{k_i}$.

Each interval in each dimension is assigned a $k_i$ bit index. **Each range represents an atomic subscription** and is assigned with a unique label. The label of each range is the concatenation of binary string of interval index in each dimension. The size of each label is $\sum_{1}^{d} k_i$. Let *k* be the average value of $k_i$, the size of each label can be denoted as *k\*d*.

Since there are $N_r = 2^{kd}$ ranges in content space, we need at least *kd* bits to assure each range label is unique. Therefore, our encoding scheme for range label is the most concise encoding and it can reduce the subscription message traffic.

For example, Figure 2 is content space with *d=2* and $k_1=k_2=5$. Each dimension index set is encoded as binary number set {000,001,010,011,100}. The content space is partition into 25 ranges, i.e. 25 atomic subscriptions.

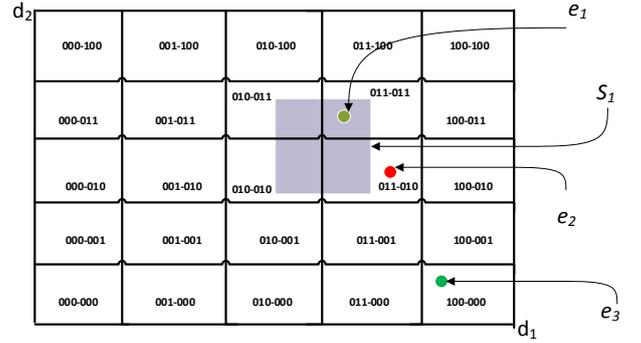

**Figure 2:** Sample events and subscription in a two dimensional space

### 3.3 Event and Subscription Representation in DLS

**Event Representation:** As discussed above, every event is a point in the content space that is represented as $e = <v_1, v_2, ..., v_d>$. Each event must belong to one range (atomic subscription) in content space. Thus, the label of the corresponding range is used as the representation of this event in DLS model.

Therefore, the vector representation is reduced to a single label. Figure 2 shows three sample events in the two dimensional space, *e1* is represented as *DLS(e1)='011-011'*, *e2* as *DLS(e2)='011-010'*, *e3* as *DLS(e3)='100-000'*.

**Subscription Representation:** To represent a subscription in the partitioned space, the first step is to augment the subscription rectangle to a rectangle that is composed of cells in the partitioned space. The resulting rectangle from the subscription augmentation step is the minimum bounding rectangle of the original subscription. This minimum bounding rectangle consists of all cells in the partitioned space that have intersection with the subscription.

Therefore, a subscription represented as a conjunction of Boolean predicates is reduced to **a set of atomic subscriptions.**

As shown in Figure 3, subscription $S_1$ is represented as $DLS(S_1)=\{'010\text{-}011', '011\text{-}011', '010\text{-}010', '011\text{-}010'\}$.

This augmentation covers some parts of the content space that are not requested by the subscription and **may result in false positives in event dissemination**. The amount of resulting false positives is directly related to the space partitioning granularity. The finer granules and smaller partitions, fewer false positives.

However, finer granules will result in a significant increase of numbers in one subscription label set. We need an efficient data structure to store label set. Based on this data structure, we can design scalable subscription aggregation algorithms and event matching algorithm. We choose Bloom filters to solve this challenge.

### 3.2 Bloom Filters

A Bloom filter is a time-efficient and space-efficient data structure with false positives, proposed by Burton Howard Bloom in 1970 [22]. It provides a way to represent and store a set or a list of items, and query whether an element is included in a set, which returns either "possibly in set" or "definitely not in set" as the result. A Bloom filter is composed of a bit vector in settled length and several hash functions. It has a strong space and time advantage over other data structures for representing sets, because it requires one bit (or several bits in counting Bloom filters) for per element in the bit vector rather than saving elements themselves.

There exists false positive rates (FPR) in Bloom filters, which means that an element, which is not in the set in reality, is wrongly evaluated as a member in this set after querying the Bloom filter. With more elements added into a Bloom filter, the false positive rate $p$ will increase. $p$ can be calculated by a formula as follows [23]:

$$p = (1 - (1 - \frac{1}{m})^{kn})^k \qquad (1)$$

In general, there is an asymptotic approximation format of Equation (1), as follows:

$$p = (1 - e^{-kn/m})^k \qquad (2)$$

In order to reduce the need for possibly computing a large number of different hash functions, the authors of [24] have shown that only two hash functions are necessary to effectively implement a Bloom filter without any loss in the asymptotic false positive probability.

There are some important variations to improve Bloom filters (which are also called "standard Bloom filters" to distinguish from variations), like counting Bloom filters [21], Dynamic Bloom filters [25], and so on.

In a counting Bloom filter, there is a counter rather than a bit in each filter entry. When adding an element, counters that are mapped by hash values will be incremented by one; when removing an element, the values will be decremented by one. As for an element query, the element is in the set if all of the corresponding counter values are non-zero. For most applications, 4 bits per counter can meet the needs [21]. Thus, counting Bloom filters can implement elements deletion.

**Definition 1.** We define $m$ as the length of the bit vector in a standard Bloom filter and the number of counters in a counting Bloom filter; $n$ is the number of elements added into the Bloom filter; and $k_h$ is the number of hash functions.

## 4 DLS Solution

### 4.1 Overview

The structure of a broker node is depicted in Figure 3. The data structures are shown in Algorithm 1.

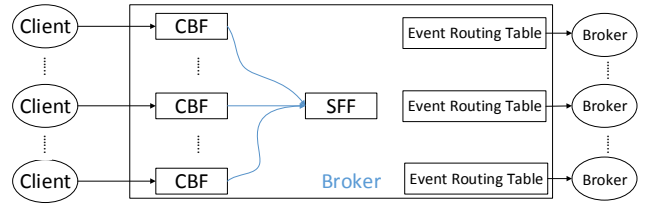

**Figure 3:** Structure of a broker

Usually one broker serves many clients and connects with a few brokers. A broker needs to maintain a subscription set for each subscriber since it depends on this knowledge to decide if an event shall be routed to this connection. Therefore, each subscriber has one CBF in its broker to maintain its subscriptions status. Since the subscription table for each client is usually small and the number of clients might be large, there are usually many small CBFs in brokers.

There is a special CBF named **SFF** (**Subscription** Aggregation **Filtering & Forwarding** Table) in each broker to represent subscriptions from all its served clients. There is one **ERT** (**Event Routing Table**) for each broker connected to this broker, as shown in Figure 3. The ERT is a CBF storing forwarded subscriptions from this connection, as shown in Figure 4.

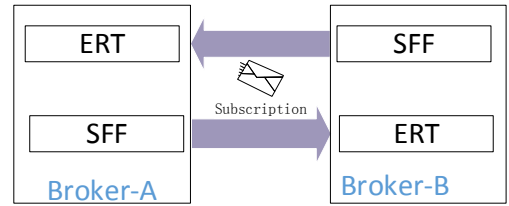

**Figure 4:** Connection between two brokers

The routing table is usually large since the SFF contains the subscriptions from thousands of clients. A broker has a routing table for each connected broker. Normally one broker only connects to a few brokers. Therefore, there are usually a few large CBFs for routing tables.

**Algorithm 1** Data Structure

1: ClientCBFs: an array with elements being CBF. The cbf is a CBF for the connection of incoming subscription message.
2: SFF: a CBF store all subscriptions in this broker node.
3: SubscriptionLabels: a set of labels to represent a subscription.
4: **ForwardingLabels**: a queue of labels not in this broker, initially ∅.
5: UnsubscriptionLabels: a set of labels to represent a unsubscription.
6: **ForwardingUnsubLabels**: a queue of labels, initially ∅.
7: Event: a label.
8: RoutingTables: an array with elements being RoutingTable, which is CBF.
9: **ForwardingEvents**: a queue of events. Each connection has a queue of forwarding events.

In content-based publish/subscribe network, the broker need to handle three type message: subscription, event, unsubscription.

The format of a subscription in DLS is a set of labels. Each label occupies $k*d$ bits. Therefore, the size of subscription message is $N_s*k*d$, where $N_s$ is the number of ranges in this subscription, i.e., the number of labels for this subscription.

When a client sends a label set (subscription) to its serving broker, the broker updates each label into the client's CBF. The repeated subscription label only results in counter increase in CBF. New labels will be checked with SFF to determine if this label is new in this broker. New labels in this broker are pushed into a queue of labels for forwarding to connected brokers. The algorithm is shown in Algorithm 2.

Unsubscriptions serves to cancel previous subscriptions. Unsubscription messages have the similar format as subscription messages. The algorithm is shown in Algorithm 3.

Each event is represented as a label in DLS model. When a client publishes an event, it sends the event to the connected broker(s). The serving broker checks each client CBF to decide if forwarding the event to the connected clients. At the same time, query routing table to decide if forwarding the event to the connected brokers.

### 4.2 Subscription Aggregation

Our subscription insertion and aggregation algorithm is shown in Algorithm 2.

The input parameter *cbf* is the CBF for the connection of incoming subscription message. SubscriptionLabels is the incoming subscription message. The labels in SubscriptionLabels will be stored into *cbf*. If the active subscription set does not contain the label, the label also needs to be inserted into SFF.

**Algorithm 2** Subscription Insertion and Aggregation Algorithm.

**Input** SubscriptionLabels, *cbf*
**Output** ForwardingLabels
1: ForwardingLabels ← ∅
2: **for all** label *l* in SubscriptionLabels **do**
3:    *cbf*.add(*l*);
4:    **if** SFF.query(*l*) = 0 **then** // new label
5:      ForwardingLabels.push(*l*);
6:    **end if**
7:    SFF.add(*l*);
8: **end for**

**Notes:** *cbf*.query(*l*), return the minimum value of retrieved $k_h$ values from CBF vector.

**Lemma 1** The running time of Algorithm 2 is $O(1)$ and its worst case time complexity is $O(N_r)$.

**Proof**:

The Bloom filters operation takes $O(1)$ time.

Since there are at most three Bloom filters operations and one push queue operation at each iteration, the algorithm terminates at most |SubscriptionLabels| (the label set size of one subscription) iterations.

For most subscriptions, the size of subscription is small. Therefore, the average time complexity is $O(1)$.

The worst case is that the label set size of some special subscription can be very large. The upper bound is $N_r$. Therefore, the worst case time complexity is $O(N_r)$.

Our subscription deletion algorithm is shown in Algorithm 3.

**Algorithm 3** Subscription Deletion Algorithm.

**Input** UnsubscriptionLabels, *cbf*
**Output** ForwardingUnsubLabels
1: ForwardingUnsubLabels ← ∅
2: **for all** label *l* in ForwardingUnsubLabels **do**
3:    **if** SFF.query(*l*) = 1 **then**
4:      ForwardingUnsubLabels. push(*l*);
5:    **end if**
6:    **if** SFF.query(*l*) != 0 **then**
7:      SFF.delete(*l*);
8:    **end if**
9:    **if** *cbf*.query(*l*) != 0 **then**
10:     *cbf*.delete(*l*);
11:    **end if**
12: **end for**

**Lemma 2** The running time of Algorithm 3 is $O(1)$ and its worst case time complexity is $O(N_r)$.

**Proof**

The Bloom filters operation takes $O(1)$ time.

Since there are at most three Bloom filters operations and one push queue operation at each iteration, the algorithm terminates at most

|ForwardingUnsubLabels| (the label set size of one unsubscription) iterations.

For most unsubscriptions, the size of unsubscription is small. Therefore, the average time complexity is $O(1)$.

The worst case is that the label set size of some special unsubscription can be very large. The upper bound is $N_r$. Therefore, the worst-case time complexity is $O(N_r)$.

### 4.3 Event Matching

The event matching algorithm is shown in Algorithm 4.

The input parameter $rt$ is the routing table for the connected broker. The input parameter $e$ is the incoming event.

**Algorithm 4** Event Matching Algorithm
**Input** $e$, $rt$
**Output** ForwardingEvents
1:  **for all** RoutingTable $rt$ in RoutingTables **do**
2:      **if** $rt$.query($e$.label()) >= 1 **then**
3:          ForwardingEvents.push($e$);
4:      **end if**
5:  **end for**
1:  **for all** CBF $cbf$ in ClientCBFs\\{cfb(e)} **do**
2:      **if** $cbf$.query($e$.label()) >= 1 **then**
3:          ForwardingEvents.push($e$);
4:      **end if**
5:  **end for**

**Notes:** ClientCBFs\\{cfb(e)} means: except the cbf associated with this event's publisher

**Lemma 3** The running time of Algorithm 4 is $O(1)$.

**Proof**

The Bloom filters operation takes $O(1)$ time.

Since there are at most one Bloom filters operations and one push queue operation at each iteration, the algorithm terminates at most | RoutingTables | + (|ClientCBFs|-1) iterations.

| RoutingTables | is the number of broker connected to this broker node.

|ClientCBFs| is the number of clients at this broker node.

It is a constant number for a given network. Therefore, the time complexity is $O(1)$.

### 4.4 FN (False Negative) Problem

If a subscription not subsumed in active subscription set are filtered by SFF due to the false positive query of Bloom filter, we call this subscription aggregation is false positive.

**False positive subscription aggregation** make event matching false negative possible. To avoid the **false negative event matching**, the configuration of SFF and ERT in broker must obey the following two rules.

1) $SFF_{ij}$ and $ERT_{ji}$ have the same size.

If one subscription label (e.g. label $L3$ in Figure 5) make false positive query of $SFF_{ij}$, the corresponding $ERT_{ji}$ will not contain this label. While the event $e_3$ with this label $L3$ arrives $B_j$, $B_j$ will query $ERT_{ji}$ to determine if route this event to Bi. To assure the event with this label can be delivered to Bi, $SFF_{ij}$ and $ERT_{ji}$ must have the same size. This trick exploits the false positive of an event matching within routing table ($ERT_{ji}$) to counteract the false positive effect of subscription aggregation within $SFF_{ij}$ Table.

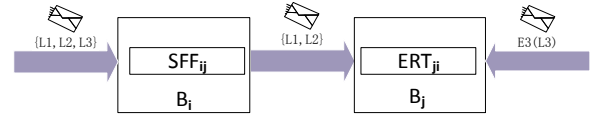

**Figure 5: False positive subscription aggregation**

2) All the SFFs along subscription propagation path have the same size.

## 5 Discussions

### 5.1 False Positive Rate

There are two sources of false positives. 1) The model mapping process from Boolean predicates to DLS model. 2) The false positive rate of Bloom filters.

Fortunately, our approximate approach provides knobs to control the accuracy by adjusting the granularity (attribution partition) of model mapping algorithm and the memory size of Bloom filters.

1) Computing with fine granules can achieve low false positive rate at the cost of high memory consumption. Since the memory capacity is known for a given system, this puts an upper bound for the size of Bloom filters. Thus, the false positive rate might raise if fine granules result in too many labels and will increase the false positive rate of Bloom filters significantly.

There is an optimal granule to achieve the lowest false positive rate. Our experimental results are shown in Section 6.

2) Our proposed DLS scheme achieves real time matching and subscription aggregation at the cost of extra false positives from Bloom filters. The parameter $m$ can be tuned by dynamic Bloom filters [25] techniques.

The network can provide real time information filter by allowing a small portion unsubscribed events to be routed to clients. The final accurate information filtering can be done by clients. If the false positive rate is in acceptable range, the overhead offloaded to clients is ignorable.

To date, we need fine tune model parameters manually. In our future work, we will develop automatic optimization techniques for model parameter.

### 5.2 Efficient Attribution Partition and Feature Selection

High dimension and fine attribution partition granules will result in the total number of labels in the content space increase significantly.

We can select representative dimensions as primary features and find attribute partition efficiently by exploiting domain knowledge. For example, doctors need monitor the body temperature data of patients. An efficient attribute partition scheme could be {Invalid, Low, Normal, LowFever, MediumFever, HighFever, Abnormal} with seven partitions. Each dimension has its own efficient partition scheme.

In our future work, we will develop the approach to optimize the DLS model parameters automatically based on the business knowledge database and mining of historical data with machine learning technology.

### 5.3 Clients Overhead Analysis

To be consistent with traditional pub/sub API, the client SDK implements the following works.
1) The DLS model mapping algorithm for events and subscriptions.
2) Exact information filtering at the client end.

However, such computation can be done very fast without requiring significant computing power if it is done in a distributed fashion by publishers and subscribers at the client end.

### 5.4 Extended Functions

**Subscriptions on Discrete Attribute:**
Each range in MICS needs an index to be organized into a B+ Tree. The range label in DLS model must be unique, but can be unordered. Therefore, it is possible to support subscriptions of discrete attributes. However, subscriptions on discrete attribute may lead to the number of ranges being increased significantly if the domain of discrete attributes is a large discrete set. This requires more memory to be assigned to Bloom filters to control false positive rate in an acceptable range.

**Supporting Multiple Businesses:**
Different applications can share the same content space and have its own DLS model parameters based on its business scenarios. Each type of application can concatenate the app ID with range labels to assure the uniqueness of each label from different applications. Thus, different applications can share the same data sources and content-based publish/subscribe infrastructure to achieve various business goals. This can avoid publishing the same event multiple times and hence reduce events traffic significantly.

## 6 EVALUATION

In this section, we evaluate the performance of the proposed method. In experiments, we compare our method with MICS [15].

### 6.1 Evaluation Environment and Data Sets

All of the experiments are conducted on a machine with a 2.6GHz Intel Xeon 6 core processor and 64 GB RAM running CentOS with Linux kernel 3.10.0−229. For each simulation run 50 GB memory is allocated for JVM.

Test data sets are generated with three parameters: data distribution, attribute partition granules and the number of dimensions. Data distribution type includes uniform distribution and Zipf distribution. We conduct simulations in 2, 3 and 4 dimensional content space.

Subscriptions and events are generated in the following way. To create a subscription, we compute the lower bound of the interval for each dimension using the selected distribution. Then we add a random interval length to the lower bound to detect the upper bound of the interval. As for events, attribute value is generated for each dimension using the selected distribution.

### 6.2 Evaluation Results

*6.2.1 Evaluate scalability with the number of subscriptions*

There are four metrics for DLS models. Since MICS do not support subscription deletion, there is no subscription deletion data for MICS.

1) Subscription insertion time.
2) Subscription forwarding traffic to evaluate Subscription aggregation effect.
3) Subscription deletion time.
4) Event matching time.

Table 1. Subscription Insertion Time under Uniform Distribution

| time ($10^{-3}$ms) | 100K | 200K | 300K | 400K | 500K |
|---|---|---|---|---|---|
| DLS(d=2, g=128) | 0.628 | 0.586 | 0.556 | 0.536 | 0.516 |
| DLS(d=3, g=32) | 0.648 | 0.623 | 0.688 | 0.563 | 0.594 |
| DLS(d=4, g=16) | 1.013 | 0.864 | 0.878 | 0.762 | 0.813 |
| MICS(d=2, g=128) | 1.840 | 2.215 | 2.460 | 2.695 | 2.804 |
| MICS(d=3, g=32) | 1.760 | 2.175 | 2.420 | 2.620 | 2.762 |
| MICS(d=4, g=16) | 2.660 | 3.070 | 3.027 | 3.320 | 3.626 |

Table 2. Subscription Insertion Time under Zipf Distribution

| time ($10^{-3}$ms) | 100K | 200K | 300K | 400K | 500K |
|---|---|---|---|---|---|
| DLS(d=2, g=128) | 0.308 | 0.309 | 0.310 | 0.302 | 0.317 |
| DLS(d=3, g=32) | 0.357 | 0.352 | 0.391 | 0.359 | 0.362 |
| DLS(d=4, g=16) | 0.413 | 0.451 | 0.446 | 0.436 | 0.488 |
| MICS(d=2, g=128) | 0.280 | 0.280 | 0.313 | 0.345 | 0.388 |
| MICS(d=3, g=32) | 0.150 | 0.165 | 0.173 | 0.173 | 0.188 |
| MICS(d=4, g=16) | 0.130 | 0.140 | 0.137 | 0.183 | 0.174 |

Both DLS and MICS demonstrate scalability with the increasing of the number of subscriptions. Data are shown in Table 1 and Table 2.

Table 3. Subscription forwarding traffic under Uniform Distribution

| Label/Range number | 100K | 200K | 300K | 400K | 500K |
|---|---|---|---|---|---|
| DLS(d=2,g=128) | 16345 | 16378 | 16376 | 16374 | 16378 |
| DLS(d=3, g=32) | 31206 | 32697 | 32756 | 32762 | 32760 |
| DLS(d=4, g=16) | 51217 | 62461 | 64832 | 65394 | 65493 |
| MICS(d=2, g=128) | 100000 | 200000 | 300000 | 400000 | 500000 |
| MICS(d=3, g=32) | 100000 | 200000 | 300000 | 400000 | 500000 |
| MICS(d=4, g=16) | 100000 | 200000 | 300000 | 400000 | 500000 |

Table 4. *Subscription forwarding traffic* under Zipf Distribution

| Label/Range number | 100K | 200K | 300K | 400K | 500K |
|---|---|---|---|---|---|
| DLS(d=2,g=128) | 128 | 128 | 128 | 127 | 129 |
| DLS(d=3,g=32) | 32 | 32 | 33 | 32 | 32 |
| DLS(d=4,g=16) | 16 | 16 | 17 | 16 | 16 |
| MICS(d=2,g=128) | 129 | 128 | 128 | 128 | 128 |
| MICS(d=3,g=32) | 32 | 32 | 32 | 32 | 33 |
| MICS(d=4,g=16) | 16 | 16 | 16 | 16 | 17 |

From the data in table 3 and table 4, DLS model show similar subscription aggregation effect as MICS model.

Table 5. *Subscription Deletion time* under Uniform Distribution

| time ($10^{-3}$ms) | 100K | 200K | 300K | 400K | 500K |
|---|---|---|---|---|---|
| DLS(d=2,g=128) | 1.212 | 1.158 | 1.137 | 1.122 | 1.000 |
| DLS(d=3,g=32) | 1.255 | 1.289 | 1.178 | 1.294 | 1.409 |
| DLS(d=4,g=16) | 1.540 | 1.394 | 1.532 | 1.551 | 1.282 |

Table 6. *Subscription Deletion time* under Zipf Distribution

| time ($10^{-3}$ms) | 100K | 200K | 300K | 400K | 500K |
|---|---|---|---|---|---|
| DLS(d=2,g=128) | 0.661 | 0.663 | 0.713 | 0.659 | 0.747 |
| DLS(d=3,g=32) | 0.825 | 0.807 | 0.787 | 0.808 | 0.776 |
| DLS(d=4,g=16) | 0.938 | 0.869 | 0.854 | 1.009 | 0.879 |

From data in table 5 and table 6, DLS model show scalability with subscription operation under the increasing number of subscriptions.

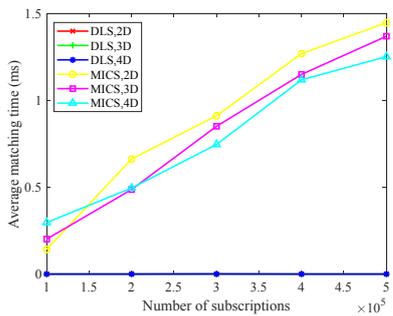

(a) Uniform Distribution

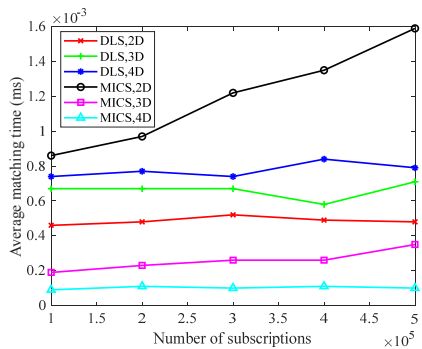

(b) Zipf Distribution

**Figure 6:** *Event matching time - various subscriptions number*

DLS demonstrates better performance of event matching time, as shown in Figure 6.

*6.2.2 Evaluate* performance under finer attribute partition granules

The selected dimension is 3. The size of subscription set is 8000 since both DLS and DLS are insensitive to the number of subscriptions – this conclusion can be drawn from both evaluation in 6.2.1 and theory analysis. The subscription size is 4 and the domain of attribute $[0, 2^{20}-1]$.

Attribute partition granules have significant impacts on the performance of MICS and less on DLS. This is because MICS organize ranges into B+ tree and the increasing number of ranges will degrade the search time. While DLS store the labels into Bloom filters, the query time is insensitive of the element number.

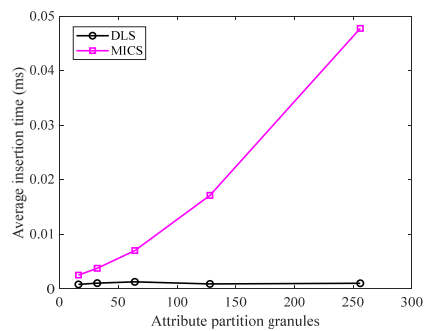

(a) Uniform Distribution

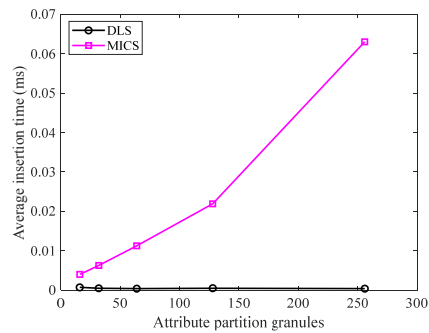

(b) Zipf Distribution

**Figure 7:** Subscription insertion time under various attribute partition granules

We compare the subscription insertion time and event matching under various attribute partition granules. Our theoretical analysis is consist with the results from experiments data.

Table 8. Subscription insertion time under various attribute partition granules

| Time ($10^{-3}$ms) | 16 | 32 | 64 | 128 | 256 |
|---|---|---|---|---|---|
| Unif，DLS | 0.791 | 1.021 | 1.250 | 0.875 | 1.000 |
| Unif，MICS | 2.5 | 3.75 | 7 | 17.125 | 47.75 |
| Zipf，DLS | 0.688 | 0.458 | 0.375 | 0.479 | 0.375 |
| Zipf，MICS | 4 | 6.25 | 11.25 | 21.875 | 63.0 |

From Figure 7 and Table 8, we can see the subscription insertion time of DLS is stable while attribute partition granules vary from 16 to 256. MICS's performance degrades significantly with the increasing of interval number.

From Figure 8 and Table 9, we can see the event matching time of DLS is stable while attribute partition granules vary from 16 to 256. MICS's performance degrades significantly with the increasing of interval number.

Table 9. Event matching time with various attribute partition granules

| Time ($10^{-3}$ms) | 16 | 32 | 64 | 128 | 256 |
|---|---|---|---|---|---|
| Unif，DLS | 0.31 | 0.35 | 0.34 | 0.405 | 0.39 |
| Unif，MICS | 36.6 | 75.1 | 84.9 | 86.1 | 88.1 |
| Zipf，DLS | 0.295 | 0.35 | 0.38 | 0.38 | 0.38 |
| Zipf，MICS | 0.1 | 0.21 | 0.36 | 0.87 | 2.51 |

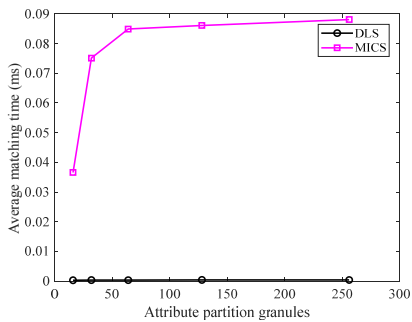

(a) Uniform Distribution

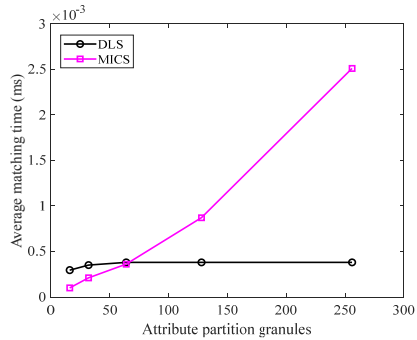

(b) Zipf Distribution

**Figure 8:** Event matching time under various attribute partition granules

6.2.*3 False positive rate.*

There are two sources of false positives. 1) The model mapping process from Boolean predicates to DLS model. 2) The false positive rate of Bloom filters.

**Finer granules impacts on false positive rate:**

Computing with fine granules can achieve low false positive rate at the cost of high memory consumption. However, a subscription will result in more labels under fine granules DLS model. While the size of Bloom filters is fixed, fine granules might also increase the false positive rate of Bloom filters significantly.

Figure 9 show experiment results under uniform distribution data set and Zipf distribution data set. 1) In red curve, the optimal granule value is 32. It means that finer granule than 32 intervals will trigger many labels in uniform distribution subscription data set and the false positive rate of Bloom filters contribute more false rate than the false rate reduction by finer granule. 2) The green curve, finer granule trigger less labels in Zipf distribution subscription data set and lead to small increasing of false positive rate of Bloom filters. Thus, total false positive rate drop as granule is finer.

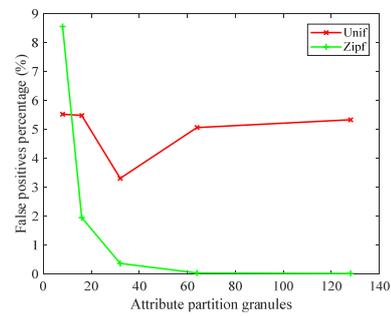

**Figure 9:** False positive rate under various granules (m=$2^{16}$bit)

**Impacts of BF size on false positive rate:**

Fix granule number as 32, evaluates the impact of Bloom filters on false positive rate of DLS model. Small BF may increase false positive rate significantly, as green curve shown in Figure 10 and Table 10.

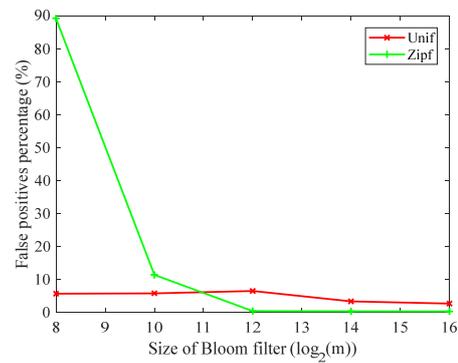

**Figure 10:** False positive rate under various size of Bloom filters

Increasing the size of Bloom filters for data set with Zipf distribution gains better results than the data set with uniform distribution.

Table 10. False positive rate under various size of Bloom filters

| m (bit) | $2^8$ | $2^{10}$ | $2^{12}$ | $2^{14}$ | $2^{16}$ |
|---|---|---|---|---|---|
| Unif(%) | 5.704 | 5.817 | 6.530 | 3.361 | 2.697 |
| Zipf(%) | 89.209 | 11.396 | 0.423 | 0.363 | 0.343 |

## 7 CONCLUSIONS

In this paper, we proposed DLS model to implement scalable subscription aggregation and real time event matching in large-scale content-based network. Our algorithms can have $O(1)$ time complexity in most cases for subscription aggregation and $O(1)$ time complexity for event matching tasks. This significant performance improvement is at the cost of memory consumption and false positive rate. Fortunately, the accuracy is controllable by adjusting model parameters. The false positive rate is controllable since finer attribute partition granules and larger Bloom filter can reduce false positive rate significantly. It can be achieved by tuning DLS model parameters.

DLS model give flexibility to application designers since they can tune attribute granules according to their business requirements. To date, model parameters tuning is manual and require domain knowledge to obtain the optimal parameters. In our future work, we will develop automatic model parameter tuning approach.

## REFERENCES

[1] P. Eugster, P. Felber, R. Guerraoui, et al, "The many faces of publish/subscribe," in *Computing Surveys (CSUR)*, 2003, pp. 114-131.

[2] A. Carzaniga, D.S. Rosenblum, and A.L. Wolf, "Design and Evaluation of a Wide-Area Event Notification Service," in *ACM Transactions on Computer Systems*, 2001, pp. 332-383

[3] Srivastava, D.: Subsumption and indexing in constraint query languages with linear arithmetic constraints. Annals of Mathematics and Artificial Intelligence 8, 315–343 (1992)

[4] A. Crespo, O. Buyukkokten, H. Garcia-Molina: Query Merging, Improving Query Subscription Processing in a Multicast Environment. IEEE Trans. Knowl. Data Eng. 15(1): pp. 174-191, (2003).

[5] Wang, Xiang, Ying Zhang, Wenjie Zhang, Xuemin Lin, and Wei Wang. "Ap-tree: Efficiently support continuous spatial-keyword queries over stream." In Data Engineering (ICDE), 2015 IEEE 31st International Conference on, pp. 1107-1118. IEEE, 2015.

[6] Margara, Alessandro, and Gianpaolo Cugola. "High-performance publish-subscribe matching using parallel hardware." IEEE Transactions on Parallel and Distributed Systems 25, no. 1 (2014): 126-135.

[7] S. Whang, H. Garcia-Molina, C. Brower, et al, "Indexing boolean expressions," Proceedings of the VLDB Endowment, vol. 2, no. 1, 2009, pp. 37-48.

[8] M. Fontoura, S. Sadanandan, J. Shanmugasundaram, et al, "Efficiently evaluating complex boolean expressions," in *SIGMOD*. ACM, 2010, pp. 3-14.

[9] Y. Zhao and J. Wu, "Towards approximate event processing in a large-scale content-based network," in *ICDCS*. IEEE, 2011, pp. 790-799.

[10] F. Fabret, H. Jacobsen, F. Llirbat, et al, "Filtering algorithms and implementation for very fast publish/subscribe systems," in *SIGMOD*. ACM, 2001, pp. 115–126.

[11] Sadoghi, Mohammad, and Hans-Arno Jacobsen. "Be-tree: an index structure to efficiently match boolean expressions over high-dimensional discrete space." In Proceedings of the 2011 ACM SIGMOD International Conference on Management of data, pp. 637-648. ACM, 2011.

[12] Shi, Ruisheng, Yang Zhang, Lina Lan, Fei Li, and Junliang Chen. "Summary instance: scalable event priority determination engine for large-scale distributed event-based system." International Journal of Distributed Sensor Networks11, no. 10 (2015): 390329.

[13] S. Qian, J. Cao, Y. Zhu, et al, "H-Tree: An Efficient Index Structure for Event Matching in Publish/Subscribe Systems," in *Ifip NETWORKING Conference*. IEEE, 2013, pp. 1-9.

[14] Z. Jerzak, C. Fetzer, "Bloom filter based routing for content-based publish/subscribe," in *DEBS*. ACM, 2008, pp. 71-81.

[15] H. Jafarpour, S. Mehrotra, N. Venkatasubramanian, et al, "**MICS**: an efficient content space representation model for publish/subscribe systems," in *DEBS*. ACM, 2009, pp. 1-12.

[16] H. Jafarpour, B. Hore, S. Mehrotra, et al, "Subscription subsumption evaluation for content-based publish/subscribe systems," in *middleware 2008*. ACM, 2008, pp. 62-81.

[17] Ouksel, A.M., Jurca, O., Podnar, I., Aberer, K.: Efficient Probabilistic Subsumption Checking for Content-Based Publish/Subscribe Systems. In: Proceedings of Middleware 2006, pp. 121–140 (2006)

[18] G. Mˉuhl. Generic constraints for content-based publish/subscribe systems. In C. Batini, F. Giunchiglia, P. Giorgini, and M. Mecella, editors, Proceedings of the 6th International Conference on Cooperative Information Systems (CoopIS '01), volume 2172 of LNCS, pages 211–225, Trento, Italy, 2001. Springer-Verlag.

[19] Li, Guoli, Shuang Hou, and Hans-Arno Jacobsen. "A unified approach to routing, covering and merging in publish/subscribe systems based on modified binary decision diagrams." Distributed Computing Systems, 2005. ICDCS 2005. Proceedings. 25th IEEE International Conference on. IEEE, 2005.

[20] E. Fidler, H. A. Jacobsen, G. Li, et al, "The padres distributed publish/subscribe system," in Feature Interactions in Telecommunications and Software Systems, 2005, pp. 12-30.

[21] L. Fan, P. Cao, J. Almeida, et al, "Summary Cache: A Scalable Wide-Area Web Cache Sharing Protocal," in *IEEE/ACM Transactions on Networking*, 2000, pp. 281-293.

[22] B. Bloom, "Space/Time Tradeoffs in Hash Coding with Allowable Errors," in *Communications of the ACM*, 1970, pp. 422-426.

[23] A. Broder, M. Mitzenmacher, "Network Applications of Bloom Filters: A Survey," *Internet Mathematics*, 2004, pp. 485-509.

[24] A. Kirsch and M. Mitzenmacher. Less hashing, same performance: Building a better bloom filter. In 14th Annual European Symposium on Algorithms, page 456-467, 2006.

[25] Guo, D., et al. "Theory and Network Applications of Dynamic Bloom Filters." IEEE INFOCOM 2006. IEEE International Conference on Computer Communications IEEE, 2006, pp:1-12.